\documentclass[aps,pre,twocolumn]{revtex4-2}
\usepackage{amssymb}
\usepackage{amsmath,bm}
\usepackage{graphicx,color}
\usepackage{bbm}
\usepackage[bottom]{footmisc}
\usepackage{multirow}
\usepackage{xcolor}
\usepackage{xpatch} 
\makeatletter   
\usepackage{xpatch} 

\newcommand{\B}[1]{{\bm{#1}}}

\newcommand{\rout}{r_\text{out}}
\newcommand{\C}[1]{{\mathcal{#1}}}

\begin{document}
\title{An intermediate phase between jammed and un-jammed amorphous solids}

\author{Yuliang Jin$^1$, Itamar Procaccia$^2$ and Tuhin Samanta$^2$}
\affiliation{$^1$Institute of Theoretical Physics, Chinese Academy of Sciences,Beijing 100190, China, $^2$Dept. of Chemical Physics, The Weizmann Institute of Science, Rehovot 76100, Israel}

\begin{abstract}
	A significant amount of attention was dedicated in recent years to the phenomenon of jamming of athermal amorphous solids by increasing the volume fraction of the microscopic constituents.
	At a critical value of the volume fraction, pressure shoots up from zero to finite values with a host of critical exponents discovered and discussed. 
	In this letter, we advance evidence for the existence of a second transition, within the jammed state of two-dimensional granular systems, that separates two phases characterized by different mechanical screening regimes.  
	Explicitly, highly packed systems are quasi-elastic with quadrupole-screening, and more loosely jammed systems exhibit anomalous mechanics with dipole screening. Evidence is given for a clear transition between these two regimes, reminiscent of the intermediate hexatic phase of crystal melting in two-dimensional crystals. 
\end{abstract}
\maketitle

The concepts of rigidity and jamming in granular systems are topics of intense studies for the community of amorphous solids \cite{96JNB,05Wya,10LN}. For athermal hard sphere systems, the transition from zero to infinite pressure upon jamming, is clearly defined in terms of a critical isostatic number of contacts $Z^*$ \cite{12KPZ,13KPUZ,14CKPUZ}. This mathematically clean phenomenology is getting somewhat blurred when temperature, or friction, are included in the relevant physics of amorphous solids. Notably, even the packing fraction in which jamming is observed is not well defined, since it can depend on details of preparation or compressing protocols \cite{10CBS}. 

The difficulty in identifying a sharp transition between solids and liquids is not particular to amorphous solids. In fact, it is well known that between perfect crystals in two dimensions and their fluid phase there exist an intervening hexatic phase in which the material exhibits intermediate properties between solid and liquid \cite{79NH}. Motivated by the celebrated theory of the hexatic phase transition we have recently investigated the anomalous mechanics that appears between elastic amorphous solids and their fluid analogs \cite{21LMMPRS,22MMPRSZ,22BMP,22KMPS,22CMP}. In this Letter we present evidence that the transition between an elastic phase and the novel (rigid) phase in which elasticity is anomalous, is in fact sharp, at least in two dimensions. We emphasize that this transition is different from the jamming or rigidity transition that has been studied in the literature \cite{96JNB,05Wya}. The different phases are characterized by different responses to non-uniform strains, such that the resulting displacement field is sensitive to different mechanical screening mechanisms, which are either  quadrupolar or dipolar elastic charges. 

The concept of screening in modern physics arises naturally in the electrostatics of continuous media, where microscopic mechanisms are available to damp the externally imposed electric field. 
A famous example of non-electric screening is the Kosterlitz–Thouless (KT) transition \cite{16Kos}, where vortex-pairs dissociate to produce a coulomb gas phase where monopoles (unbound vortices) are available to screen external fields, similar to the Debye-screening in the electrostatic analog. 
In the field of mechanics, the nucleation of structural defect in response to external loads, is the basis for theories of crystal melting \cite{78HN,02Nel}, where solids are supplemented with screening quadrupoles (dislocation pairs), hexatics with screening dipoles (unbound dislocations) and  liquids with screening monopoles (unbound disclinations). 

While thermal melting and the hexatic phase transition necessitate finite-temperature statistical mechanics, recent research on a variety of {\em athermal} systems pointed out the existence of similar transition in cellular tissue models \cite{20PYNP} and vibrated granular matter \cite{05OU,16SLMZ,22ZVJ}. 
These observations were fundamentally based on structural characteristics, including short and quasi-long range translational and orientational order, observed via correlation functions.

In contrast to these thermal or mechanically agitated examples of hexatic phases, we present here theoretical and numerical indications for the existence of a transition to a dipoles-screened phase within the {\em jammed regime} of {\em athermal} amorphous granular systems. This
finding is based on recent research, in which it was discovered that the prevalence of plastic events in amorphous solids results in screening phenomena that are akin, but richer and different, to screening effects in electrostatics \cite{21LMMPRS,22MMPRSZ,22BMP,22KMPS,22CMP}. Plastic events, which are typically quadrupoles in the displacement field, can act as screening charges. It was shown that when the density of plastic quadrupoles is low, their effect is limited to renormalizing the elastic moduli, but the structure of (linear) elasticity theory remains intact. This is analogous to dipole screening in dielectrics. On the other hand, when the nucleation cost of quadrupoles drops, the quadrupoles density
becomes high, and the nucleation of effective dipoles defined by the gradients of their density, cannot be neglected. The presence of effective dipoles changes the analytic form of the response
to strains, in ways that are in fundamental clash with standard elasticity theory. It was concluded that one needs to consider a new theory, and this
emergent theory was confirmed by comparing its predictions to results of extensive experiments and simulations \cite{21LMMPRS,22MMPRSZ,22BMP,22KMPS,22CMP}. While dipole screening was observed in both two and three dimensions, in this letter we focus on two dimensional systems in which one can demonstrate a clear transition as stated above.  

Having realized that {\em gradients of quadrupole density} act as effective dipoles, it became evident that the vast majority of experiments and numerical simulations that study the mechanics of amorphous solids are not using the best strain protocols. Indeed, in most studies researchers employ simple and pure shear, or tensile compression or extension. To expose the unusual and interesting mechanical properties of amorphous solids it is advisable to employ non-uniform strains. To this aim we have employed, in both experiments \cite{22MMPRSZ} and simulations \cite{21LMMPRS,22MMPRSZ,22KMPS}, a circular geometry in two dimensions and a spherical one in three dimensions \cite{22CMP}. In the former case we then inflate a central disk,  to observe the resulting radial component of the displacement field $d_r$. Theoretically we expect that in a phase where only quadrupole screening is dominant, an inflation $r_{\rm in}\to r_{\rm in}+d_0$ of a disk of initial Radius $r_{\rm in}$ in a system of outer radius $r_{\rm out}$ will result in radial displacement
\begin{eqnarray}
	d_r (r) = d_0 \frac{r_{\text{in}} \left(r^2-r_{\text{out}}^2\right)}{r \left(r_{\text{in}}^2-r_{\text{out}}^2\right)},\quad \text{in two dimensions}\ .
	\label{renelas}
\end{eqnarray}
On the other hand, in a phase that is governed by dipole screening we theoretically expect
the same radial component of the displacement field to obey
\begin{equation}
	d_r(r)  = d_0 \frac{ Y_1(r \, \kappa ) J_1(r_\text{out} \kappa )-J_1(r \, \kappa ) Y_1(r_\text{out} \kappa )}{Y_1( r_\text{in} \kappa ) J_1(r_\text{out} \kappa )-J_1(r_\text{in} \kappa ) Y_1(r_\text{out} \kappa )} \ .
	\label{amazing}
\end{equation}
Here $J_1$ and $Y_1$ are the circular Bessel functions of the first and second kind respectively. The parameter $\kappa$ has the dimension of inverse length and is referred to as the screening parameter. When $\kappa\to 0$ the expression (\ref{amazing}) tends to Eq.~(\ref{renelas}).
\begin{figure}
	\includegraphics[width=0.7\linewidth]{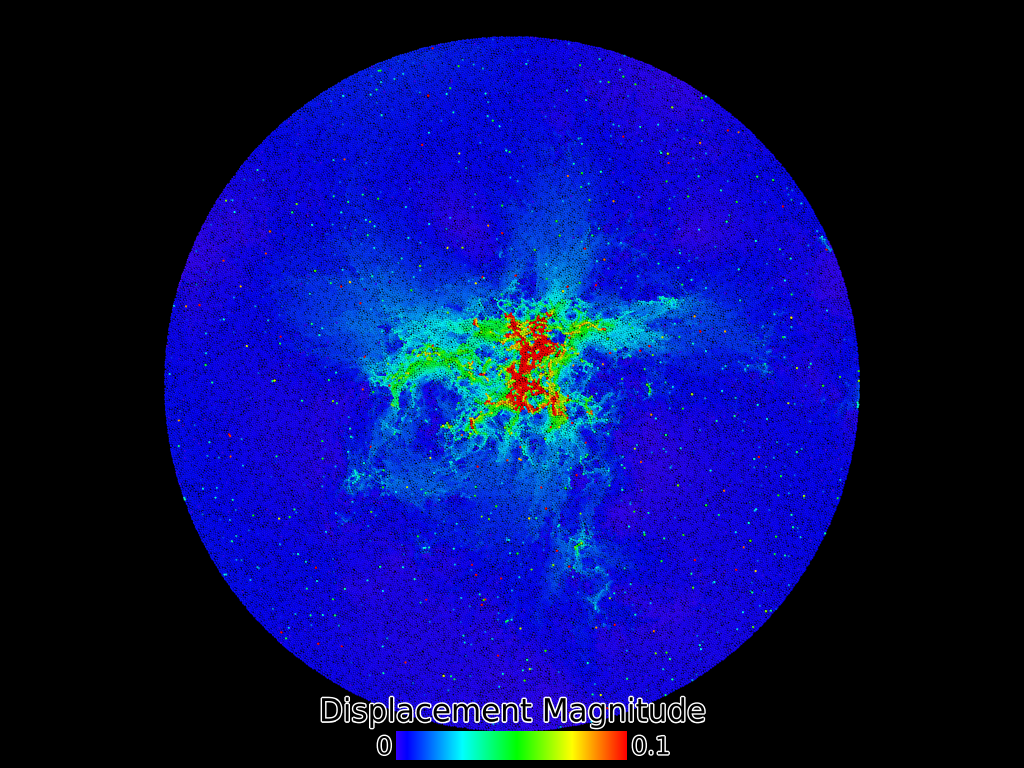}
	\includegraphics[width=0.7\linewidth]{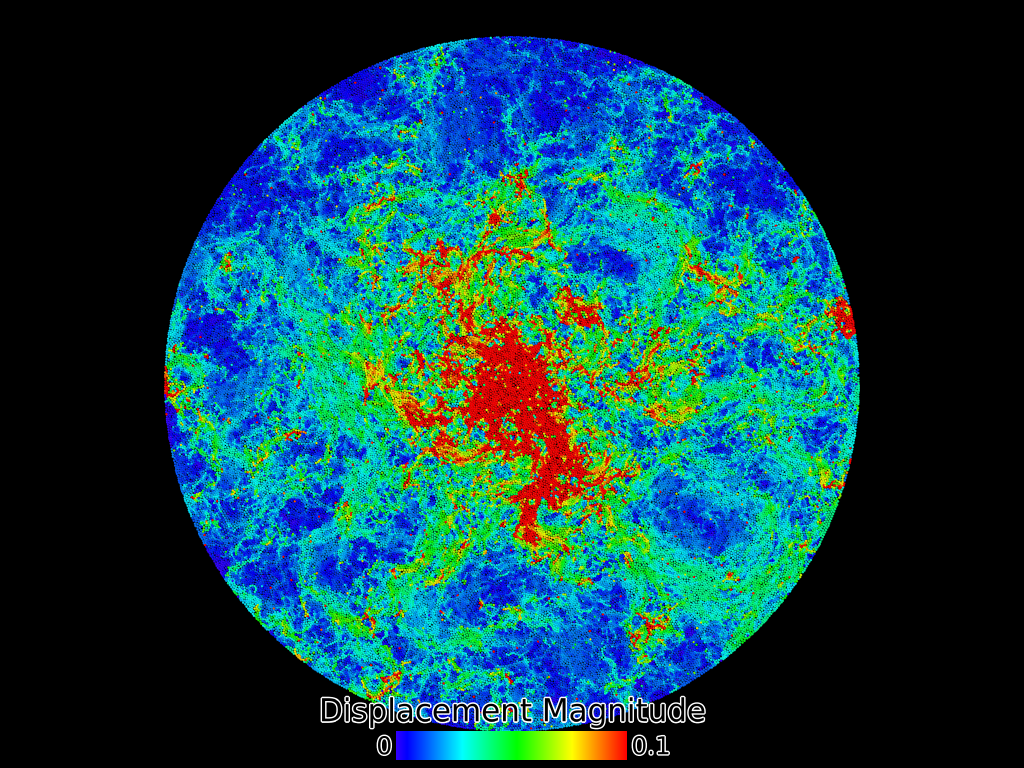}
	\caption{Maps of the magnitude of the displacement field after inflation of 25\% in the radius of the inner disk, $N=80000$. Upper panel: high pressure $P=29.35$. Lower panel: low pressure $P=0.394$}.
	\label{maps}
\end{figure}

Obviously, the difference between Eqs.~(\ref{renelas}) and (\ref{amazing}) is striking. The first
exhibits a monotonous decay of an outward displacement until it vanishes at the outer boundary,
whereas the latter allows oscillations, and even negative (inward pointing) displacements although the imposed inflation points outwards. The main question that we raise here is whether there exists
a clear transition, as a function of an intensive parameter in a given athermal amorphous system,
separating material phases in which the mechanical response tends to jump from Eq.~(\ref{renelas}) to Eq.~(\ref{amazing}) with a finite value of $\kappa$. We show next that in 2-dimensions the answer is affirmative, the intensive parameter for a granular jammed system is the pressure, and the transition is indeed clear.  

To demonstrate the transition we investigate frictionless assemblies of small disks that are at mechanical equilibrium,  prepared with a desired target pressure $P$ and confined in a circular two-dimensional area with a fixed outer circular wall. Open source codes (LAMMPS \cite{95Pli}) are used to perform the simulations. Every simulation begins with a configuration of $N=80000$ bi-disperse disks of mass $m=1$, placed randomly in a circular area with a radius $\rout=172$  in dimensionless units.  Half of the small spheres have a radius $R_1 =0.45$ and the other half a radius $R_2=0.65$.  One larger disk is not placed randomly, but rather fixed to the center of coordinates. To reach a desired pressure we begin with a chosen packing fraction and the system is relaxed to mechanical equilibrium by solving Newton's second law of motion with damping. This process is carried out until the desired target pressure is reached and forces are minimized to values smaller than $10^{-6}$.
The normal contact force is Hertzian with force constant $k_n=2\times 10^5$, following the Discrete Element Method of Ref.~\cite{79CS}. The tangential contact force is zero as the system is frictionless.

\begin{figure}
	\includegraphics[width=0.8\linewidth]{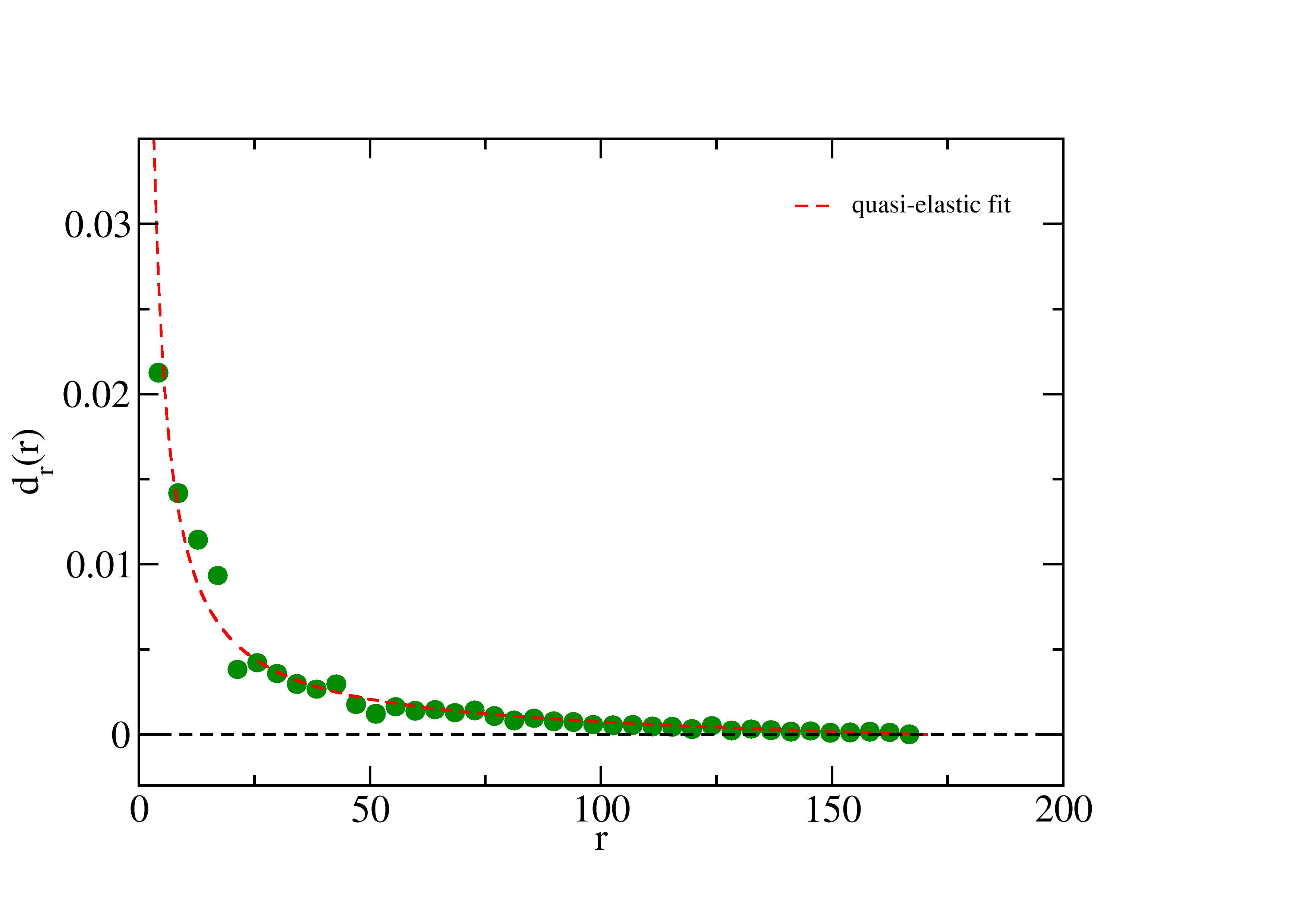}
	\includegraphics[width=0.8\linewidth]{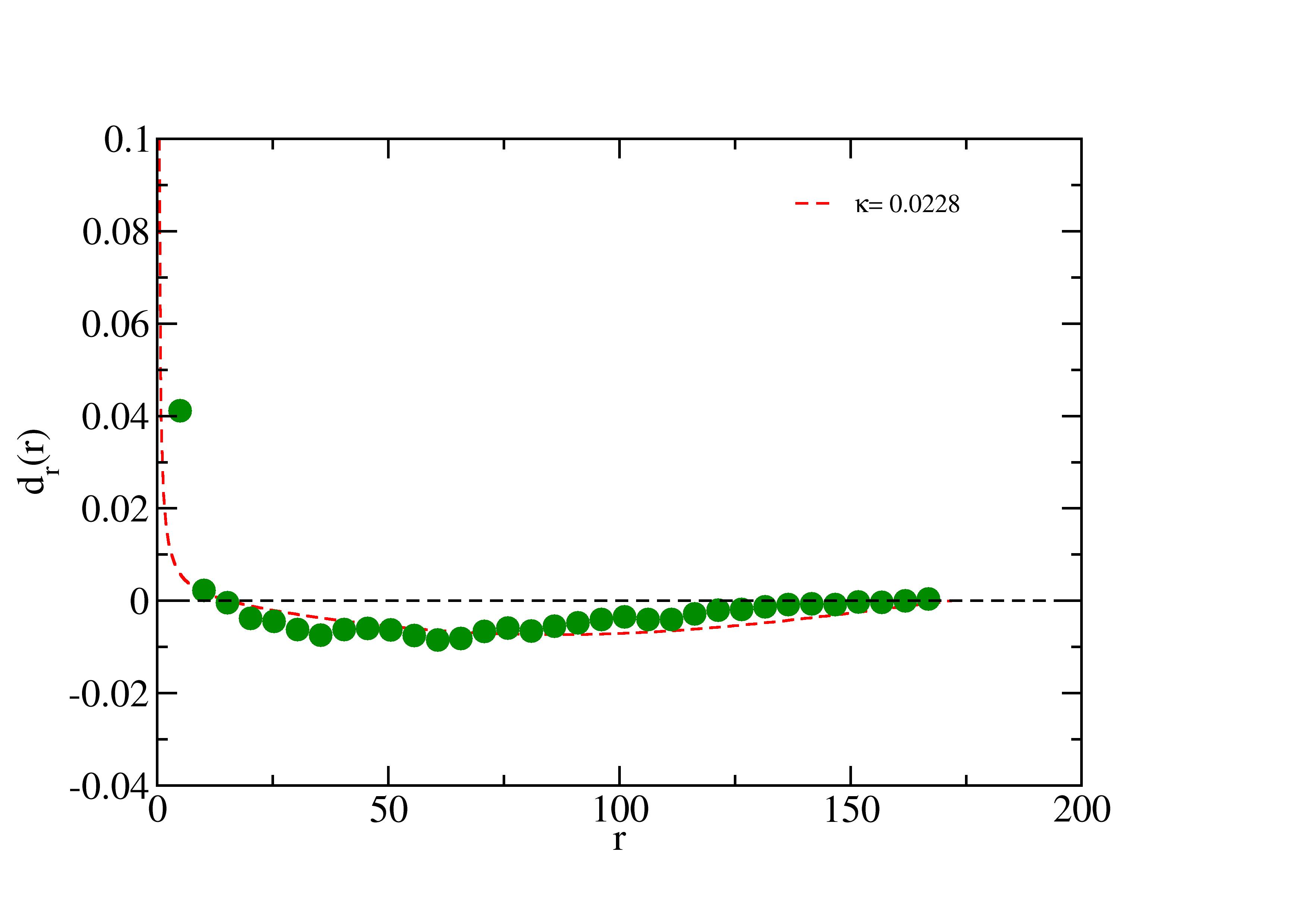}
	\caption{Green dots: the radial component of the displacement field that corresponds to the data in Fig.~\ref{maps}, averaged over the angles. Dashed line is the fit to theory. Upper panel: high pressure $P=29.35$. Lower panel: low pressure $P=0.394$, $\kappa=0.023$ }
	\label{radial}
\end{figure}

After achieving a mechanically stable configuration at a target pressure, we inflate the central disk by 25\%. The displacement field is denoted $\B d(r,\theta)$ and the radial component is obtained as an angle average, $d_r(r)\equiv (2\pi)^{-1} \oint_{0}^{2\pi} \B d\cdot \hat{r} d\theta$ where $\hat r\equiv \B r/r$. The displacement field exhibits qualitatively different appearance at high and low pressures as exemplified in Fig.~\ref{maps}.
At high pressures the displacement field is centered around the inflated disk as is expected from Eq.~(\ref{renelas}). In contrast, at low pressure the displacement field is spread out throughout the system, in correspondence with Eq.~(\ref{amazing}).
A quantitative comparison is provided by plotting the radial component $d_r(r)$, cf. Fig.~\ref{radial}.
While the upper panel shows the typical decay of an elastic solution, the lower panel presents {\em negative} radial displacement  that result from screening and the Bessel functions in Eq.~(\ref{amazing}).

\begin{figure}
	\includegraphics[width=1.0\linewidth]{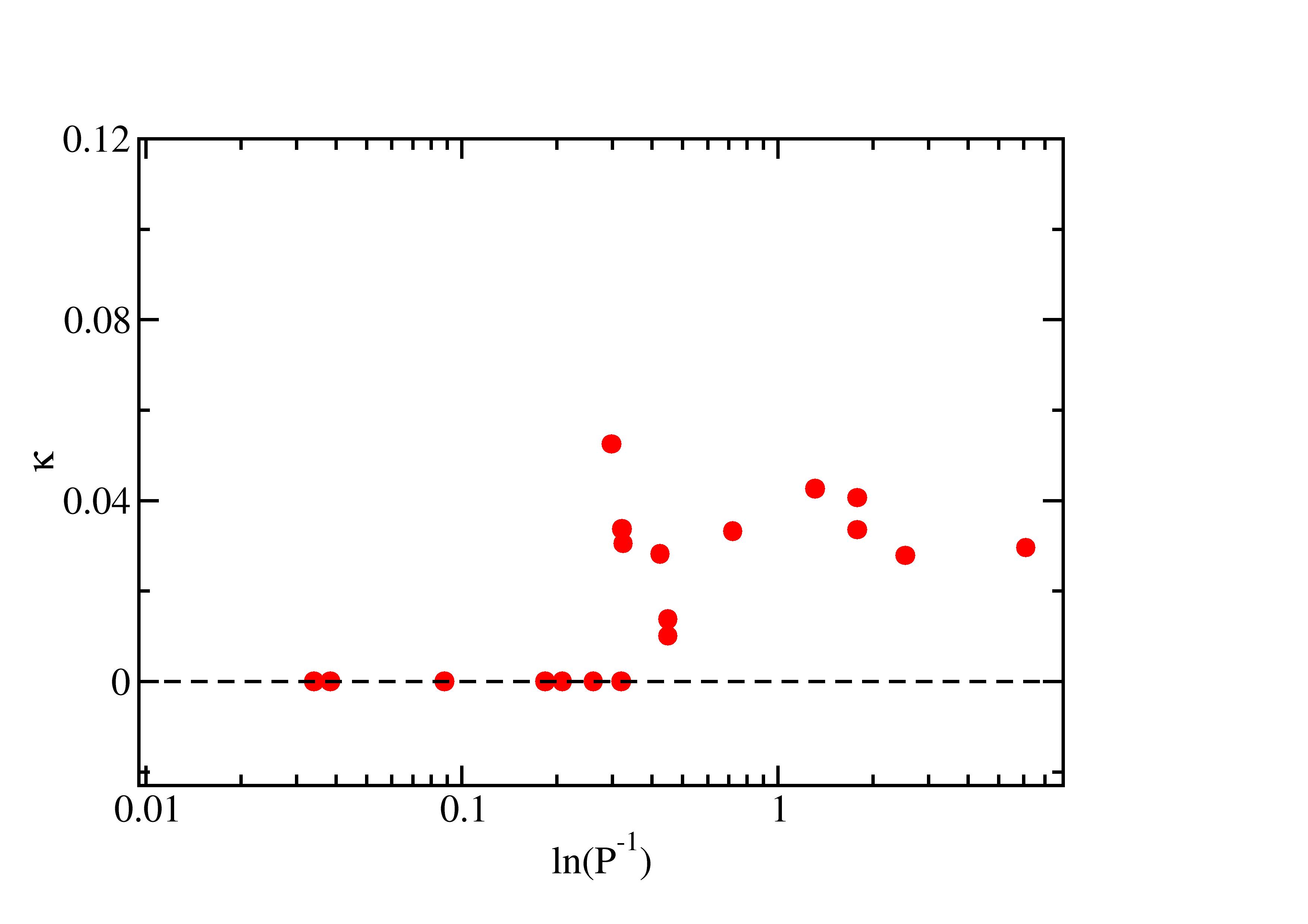}
	\caption{The screening parameter $\kappa$ as a function of the logarithm of the inverse pressure. A transition between material phases with quasi-elastic response and with anomalous response is clearly observed.}
	\label{kapP}
\end{figure}

The simulations indicate a clear transition from quasi-elastic to anomalous response. The best way to demonstrate the transition is to measure the screening parameter $\kappa$ as a function of the pressure. In Fig.~\ref{kapP} we present the measured screening parameter as a function of $\ln(P^{-1})$.
The screening parameter was measured in two independent ways. In one we simply fitted the measured radial component of the displacement field to Eq.~(\ref{amazing}), see for example Fig.~\ref{radial} lower panel. The second method relied on a direct measurement of the presence of dipoles.   This will connect the observed transition to the well known hexatic phase transition which appears in two-dimensional melting. It should be stressed that in the present context the existence of dipoles, or in fact of a dipole density $\B{\C P}(r,\theta)$, does not refer to the material structure, but rather to the presence of dipoles in the displacement field. To this aim we refer to the theory presented in Ref.~\cite{22BMP}. It was shown that the dipole density can be measured directly by the following line integral:
\begin{equation}
	\oint _{\partial \Omega }\B{\C P}(r,\theta)\cdot \mathbf{n}	\, \text{dl}= \oint _{\partial \Omega } (\nabla ( \nabla \cdot \B d)) \cdot \mathbf{n}\, \text{dl}   \ .
	\label{P}
\end{equation}
The line integral can be taken around any closed loop. Due to obvious conservation laws we expect that when the loop encircles the whole system, the net dipole included should be zero, whereas with a loop enclosing any part of the system the integral will not vanish {\bf only when there is dipole density in the enclosed area $\Omega$}. 
In the case of our circular systems with radial symmetry both sides of this equation
can be evaluated analytically. The final result is ~\cite{22BMP}
\begin{equation}
	\oint _{\partial \Omega }\B{\C P}(r,\theta)\cdot \mathbf{n}	\, \text{dl}=-\kappa^2	\oint _{\partial \Omega }\B d(r,\theta)\cdot \mathbf{n}	\, \text{dl} \ .
	\label{dipole1}
\end{equation}

\begin{figure}
	\vskip 1.2 cm
	\includegraphics[width=1.0\linewidth]{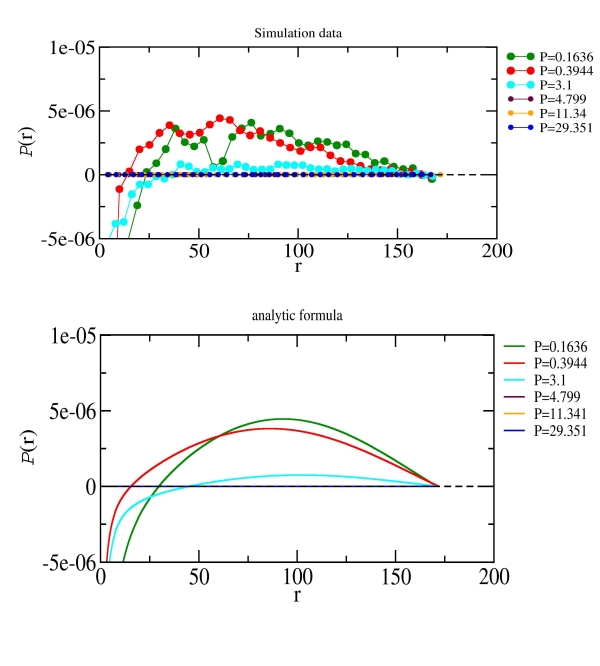}
	\caption{The dipole density included in a radius of size r, as a function of r, for the system with $N=80000$, for different pressures. Upper panel: calculation using the simulation data. Lower panel: calculation based on the analytic formulae (\ref{dipole1}) with $d(r)$ taken from Eqs. (1) or (2) (with the measured value of $\kappa$).}
	\label{dipoles}
\end{figure}

In Fig.~\ref{dipoles} we present the function $\C P(r)$ computed as a function of the radius $r$ of the loop integral for our system with $N=80000$ disks. 
The function $\C P(r)$ is computed in two ways. In the upper panel the simulation data was used according to Eq.~(\ref{dipole1}), whereas in the lower panel Eq.~(\ref{dipole1}) was applied to the analytic formulae Eqs.~(\ref{renelas}) or (\ref{amazing}) (with the measured value of $\kappa$). At pressures with quasi-elastic response, $\C P(r)$ vanishes for every value of $r$. In the anomalous regime $\C P(r)$ is not zero, in very good agreement between the two methods of computation, showing that the transition is tightly associated with the appearance of dipole densities in the displacement field at these conditions. Finally, the value of the screening parameter $\kappa$ was determined by taking the ratio of the two integrals in Eq.~(\ref{dipole1}) , giving us $\kappa^2$. 

The values of the screening parameter $\kappa$ as a function of inverse pressure is shown in linear-log scales in Fig.~\ref{kapP}. The values shown were obtained as an average between the two methods of measurement. For pressure $P\ge 3.5\pm 0.3$ the response is quasi-elastic with $\kappa=0$. For pressure $P\le 3.5\pm 0.3$ the response is anomalous. The scatter in the values of $\kappa$  in the anomalous regime is typical to the considerable sample-to-sample fluctuations in the values of the screening parameter. We should note that once the screening parameter differs from zero it appears quite independent of pressure. 

\begin{figure}
	\includegraphics[width=1.0\linewidth]{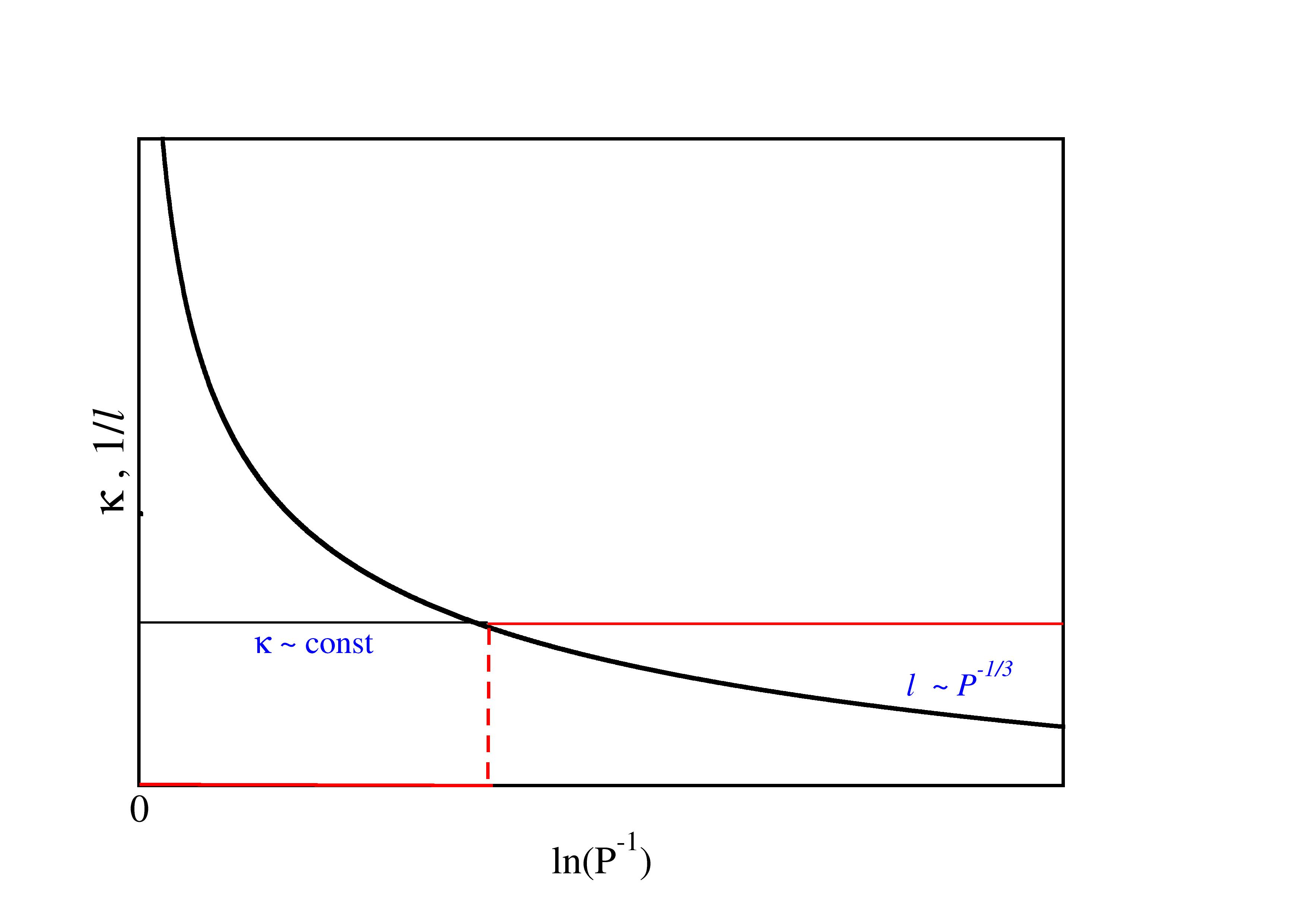}
	\caption{A sketch of the expected transition in the observed value of the screening parameter $\kappa$}.
	\label{sketch}
\end{figure}

To understand the transition and the apparent constancy of $\kappa$ as a function of pressure, we need to determine when we can expect an avalanche of plastic events that can span a region of size $\kappa^{-1}$. Start with estimating the maximal size of a blob that can become unstable and go through a plastic deformation. In our amorphous configurations of disks interacting via Hertzian forces.
the pressure depends on excess coordination number $\Delta Z\equiv Z-Z^*$ according to \cite{10LN,18BMHM}
\begin{equation}
	p\sim (\phi-\phi_J)^{3/2} \sim \Delta Z^3 \ ,
\end{equation}
where $Z^*=4$ is the coordination number at jamming. When $\Delta Z=0$, breaking any contact will render the system unstable. On the other hand, when $\Delta Z>0$ one can afford breaking more than one bond, in fact one can break a whole circumference of bonds of length $\ell^{d-1}$ as long as \cite{05WNW},
\begin{equation}
	Z\ell^{d-1} = \Delta Z \ell^d \ .
\end{equation}
If the region is smaller than this $\ell$, it is unstable to such breaking, and if larger, the region is always stable and rigid. 
We then interpret this length as the maximal blob size that can participate in an avalanche of plastic events. This length scale depends on pressure like
\begin{equation}
	\ell \sim \frac{Z}{\Delta Z} \sim p^{-1/3} \ .
\end{equation}
This is represented by the black line in Fig.~\ref{sketch}.

Next the question is what is the pressure dependence of $\kappa$. From Ref.~\cite{22CMP} one reads
\begin{equation}
	\frac{\mu_1^2}{\mu_2 \left(\lambda + 2\mu\right)}  =\kappa^2  \ .
\end{equation}
Here $\mu_1$ and $\mu_2$ are new moduli associated with the quadrupole and dipole terms in the energy function. The combination of Lame' coefficients $(\lambda + 2\mu)=B+G$,
where $B$ and $G$ are the bulk and shear moduli respectively.  This combination
is dominated by $B\sim (\phi-\phi_J)^{\alpha-2}$ because
$G\sim (\phi-\phi_J)^{\alpha-3/2}$ with $\alpha=5/2$. The coefficients $\mu_1$ and $\mu_2$ are second derivatives of the energy with respect to strain, either directly or through the quadrupole field $Q$. Thus they are both expected to scale like $B$. Therefore $\kappa$ should be independent of pressure. However, $\kappa$ can exist only when a blob of the order of $\kappa^{-1}$ exceeds the scale $\ell$. Accordingly we predict that the {\em observed} value of $\kappa$ will be zero for high pressures and constant for small pressures, with a jump when
$\kappa\approx \ell^{-1}$. This is the red line in Fig.~\ref{sketch}. Here we propose that the sketch 
presented in Fig.~\ref{sketch} rationalizes the numerical results shown in Fig.~\ref{kapP}. We should note however that the pressure where the transition is observed can depend on the magnitude of the inflation at the central disk and on the microscopic properties of the amorphous material. 

{\bf Acknowledgments}: IP thanks Michael Moshe for very useful discussions. This work has been supported in part by the the joint grant between the Israel Science Foundation and the National Science Foundation of China, and by the Minerva Foundation, Munich, Germany. 

\bibliography{ALL.anomalous}
\end{document}